\documentclass[]{spie}  %>>> use for US letter paper
\usepackage[]{graphicx}

\title{An update of the on-sky performance of the Layer-Oriented wave-front
sensor for MAD}
\author{Carmelo Arcidiacono\supit{a,b} Matteo Lombini\supit{c} Alessia Moretti\supit{b} Roberto Ragazzoni\supit{b} Jacopo Farinato\supit{b} Renato Falomo\supit{b} and Marco Gullieuszik\supit{b} and Giampaolo Piotto\supit{d}
\skiplinehalf
\supit{a}INAF-Osservatorio Astrofisico di Arcetri, Largo Enrico Fermi, 5, Firenze, Italy; \\
\supit{b}INAF-Osservatorio Astronomico di Padova, Vicolo dell'Osservatorio, 5, Padova, Italy; \\
\supit{c}INAF-Osservatorio Astronomico di Bologna, Via Ranzani, 1, Bologna, Italy; \\
\supit{d}Universit\`a degli studi di Padova, Vicolo dell'Osservatorio, 2, Padova, Italy \\
}

\authorinfo{Further author information: (Send correspondence to Carmelo Arcidiacono)\\C.A: E-mail: carmelo@arcetri.astro.it, Telephone: +39 055 2752 293}
\newcommand\aap{{A\&A}}%
\newcommand\pasp{{PASP}}%
\begin{document}
  \def\arcmin{$^{\prime}$}
  \def\arcsec{$^{\prime\prime}$}
  \def\arcdeg{$^{\rm o}$}
  \def\Ib{{\bf I} (0.85 $\mu${\em m})$\;$}
  \def\Jb{{\bf J} (1.25 $\mu${\em m})$\;$}
  \def\Hb{{\bf H} (1.65 $\mu${\em m})$\;$}
  \def\Kb{{\bf Ks} (2.12 $\mu${\em m})$\;$}
  \def\spie{Proc. of SPIE}
  \def\aap{A\&A}
  \def\pasp{PASP}
  \def\byone{1\arcmin$\times$1\arcmin}
  \def\bytwo{2\arcmin$\times$2\arcmin}
  \maketitle

%%%%%%%%%%%%%%%%%%%%%%%%%%%%%%%%%%%%%%%%%%%%%%%%%%%%%%%%%%%%% 
\begin{abstract}
The Multi-conjugate Adaptive optics Demonstrator, MAD, successfully demonstrated on sky the MCAO technique both in Layer
Oriented and Star Oriented modes. As results of the Guaranteed Time Observations in Layer Oriented mode quality astronomy
papers have been published. In this paper we concentrate on the instrumentation issues and technical aspects which stay
behind this success. 
\end{abstract}

%>>>> Include a list of keywords after the abstract 


%%%%%%%%%%%%%%%%%%%%%%%%%%%%%%%%%%%%%%%%%%%%%%%%%%%%%%%%%%%%%
\section{INTRODUCTION}
\label{sec:intro}  % \label{} allows reference to this section
The Multi-conjugate Adaptive optics Demonstrator\cite{MADFDR,marchetti05} (MAD) was an experiment devoted to demonstrate the MultiConjugate Adaptive
Optics\cite{beckers88,beckers89a} (MCAO) concept in the framework of the ESO European Extremely Large Telescope project. Aboard MAD two different
Multiconjugate Adaptive Optics techniques have been implemented: Star Oriented multi-Shack-Hartmann and Layer Oriented
(LO) multi pyramids\cite{LO1,LO2}. MAD has been designed to retrieve partially corrected adaptive optics image on a 2$\times$2 arcmin field of view
re-imaged by the CAMCAO\cite{CAMCAO1,CAMCAO2} infrared camera. The Point Spread Function (PSF) uniformity and performance are strongly depending on the guide stars
brightness, atmospheric conditions and adaptive optics control loop setup. In this paper we present the performance as they
have been measured on data taken during the Guaranteed Time Observation (GTO) nights. The Multi-Conjugate Adaptive Optics
technique in its Layer Oriented realization have been successfully demonstrated, retrieving diffraction limited resolution point
source images largely uniform on the Field of View. We will present the instrumentation issues and technical aspects which stay
behind this success, especially taking care of the limiting magnitude achieved with respect to other techniques. We take the
opportunity to discuss the lesson learned using the instrument.

MAD was mounted on the Visitor Focus on one of the Nasmyth platforms of the VLT-Melipal (UT3) in 2007 and between 21$^{\rm st}$ to 29$^{\rm st}$ September the telescope has been scheduled for 3 Technical and 6 Guaranteed Time Observations (GTO) nights in Layer Oriented mode.
In this paper we will discuss the most important technical results obtained during these nights: we briefly describe the MAD Layer Oriented WaveFront Sensor (LOWFS hereafter) in order to have the elements to discuss the way we performed diffraction limit resolution and the limitations imposed by the technical aspects. 

\section{The MAD experiment}
\label{sec:mad}  % \label{} allows reference to this section
MCAO aims to overcome the anisoplanatism problem of the single conjugate adaptive optics correction (SCAO), currently performed on different observatories, improving in this way also the angular dimension of the corrected field and, at the end, the sky coverage. In fact SCAO provides correction only on a limited angular dimension (10arcsec-20arcsec) because the atmospheric volume seen by the wave-front sensing system is limited to the direction of the reference guide star. In MCAO the wave-front sensor (WFS) senses simultaneously the wave-fronts of several guide stars. This information allows to reconstruct the three dimensional distribution of the atmospheric optical aberration. 

At least two deformable mirrors (DM) apply the MCAO correction: DM are optically conjugated to as many atmospheric layers extending in this way the correction to the Field of View (FoV) projected on them. But it is possible to use a single deformable mirror to perform the correction of the ground, and most turbulent, layer realizing in this way the so called Ground Layer Adaptive Optics (GLAO), which provides a less efficient correction but can be applied on much larger FoVs (2\arcmin to 6\arcmin), using several guide stars adequately separated. In this latter case the limitation on the FoV size is given by the fact that larger the FoV the thinner is the turbulent layer size (deep) corrected by the DM: therefore the larger the FoV the smaller the correction performance in absolute terms (for example the Strehl Ratio).
We said that on MAD two different MCAO approaches have been implemented: the Star Oriented (SO) multi Shack Hartmann WFS\cite{marchetti06} and Layer Oriented\cite{ragazzoni02,vernet2005,2006SPIE.6272E..68A,2007MmSAI..78..708A} (LO) multi pyramid {WFS}. Both sensors look for reference stars on the central 2\arcmin FoV: the slopes measurements derived by the WFS are the input for the wavefront computer to compute the voltages for the two deformable mirrors optically conjugated to 0 and 8.5 km far from telescope pupil. Wave-front sensing is performed in visible band while the imager instrument is the CAMCAO infrared camera, which was built by the Universidad de Lisboa and was mounted on the corrected focal plane of MAD: CAMCAO is a high resolution, wide Field of view Near-InfraRed (NIR) camera, that uses the 2k$\times$2k HgCdTe HAWAII-2 infrared detector from Rockwell Scientific corresponding to 57$\times$57arcesc (0.028\arcsec/px pixelscale). CAMCAO operates in
the near infrared region between 1.0$\mu$m and 2.5 $\mu$m wavelength and using a filter wheel with J, H,
Ks, K-continuum and Br$\gamma$ filters.
  \begin{figure}
   \begin{center}
   \begin{tabular}{c}
   \includegraphics[height=6cm]{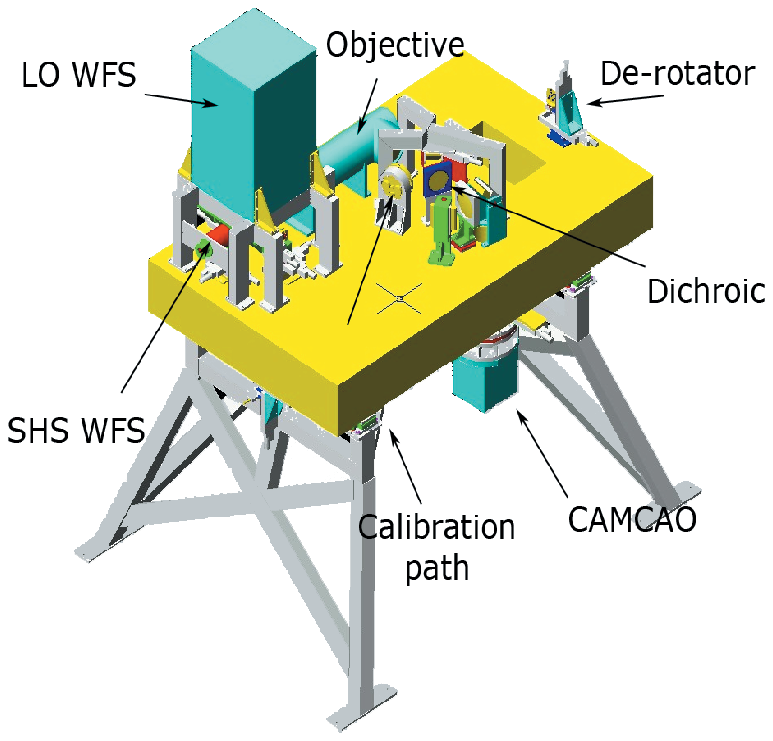}\includegraphics[height=6cm]{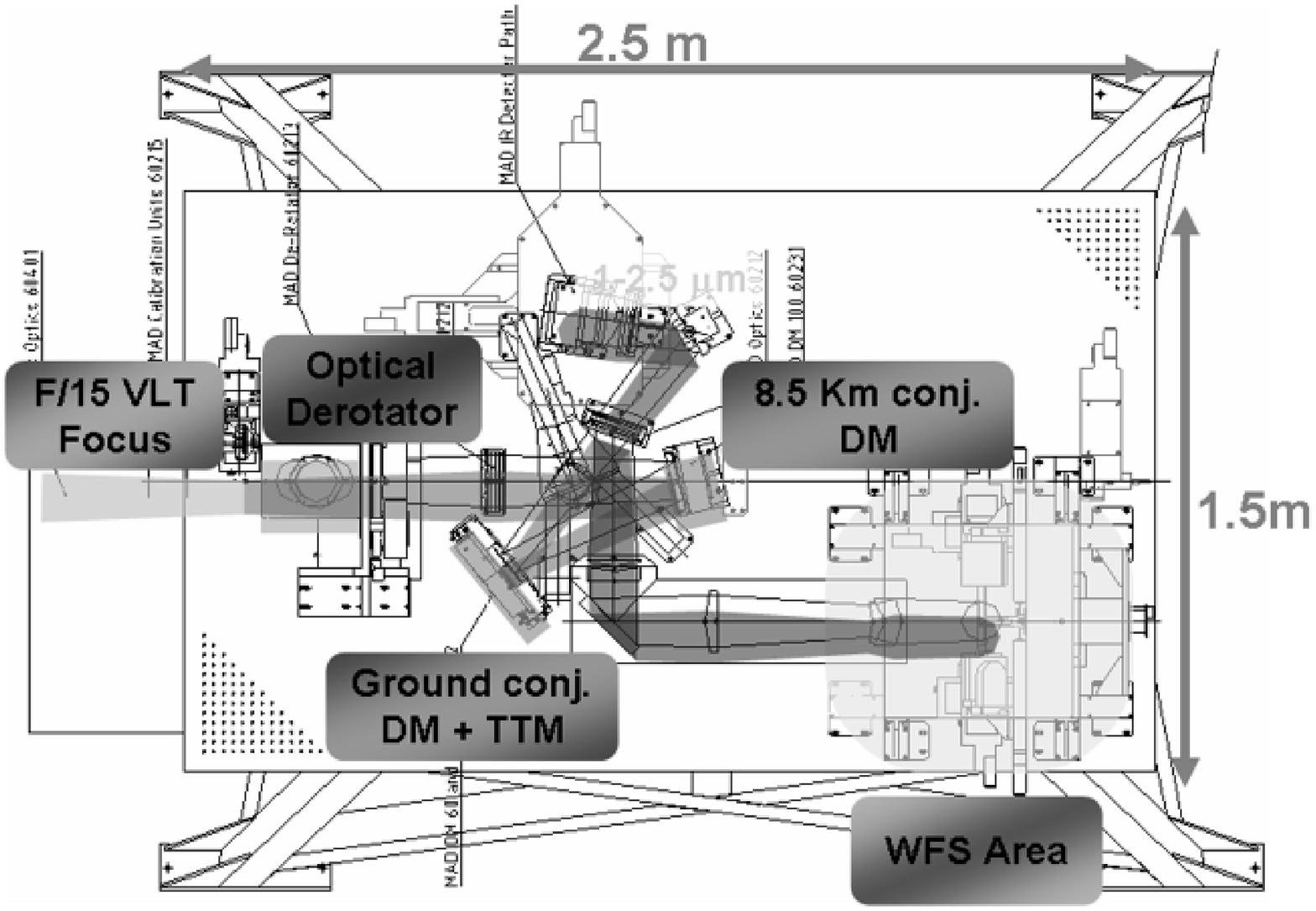}
   \end{tabular}
   \end{center}
   \caption[fig1] 
   { \label{fig:fig1} \label{fig:madbench}
The CAD view on the left presents the MAD bench and the instrument main 
components of the mechanical design concept: the beam coming from the tertiary mirror is
de-rotated and passes through the collimator. Then it is reflected by the two Deformable Mirrors (DMs). The InfraRed (IR) light is transmitted to the
camera while the visible light is reflected by the dichroic toward
the WFS objectives. The Multi Shack-Hartmann WFS is located below the LOWFS (the elongated box
on the left side). On the right a top view.}
   \end{figure} 

The MAD--bench is not fixed to the VLT--Nasmyth adapter rotator flange, then the pupil rotates with the field: an optical de--rotator at the entrance of the adaptive--system rotates both. An error in the control software of the rotator was identified and corrected during the commissioning of the instrument, unfortunately after the night-time assigned the LOWFS.

The adaptive system is illuminated by re--imaging optics collimating the F/15 input beam in order to re--image the telescope pupil on the ground layer bimorph deformable mirror and conjugating the second one to 8.5km (each deformable mirror is controlled by 60 actuators).
A dichroic transmits the IR light to the CAMCAO scientific camera while the
visible light is reflected toward to the WFS path. The MAD bench common optics retrieves to the WFS a flat telecentric F/20 input beam.

\subsection{The Layer Oriented WFS}
  \begin{figure}
   	\includegraphics[height=2.5in]{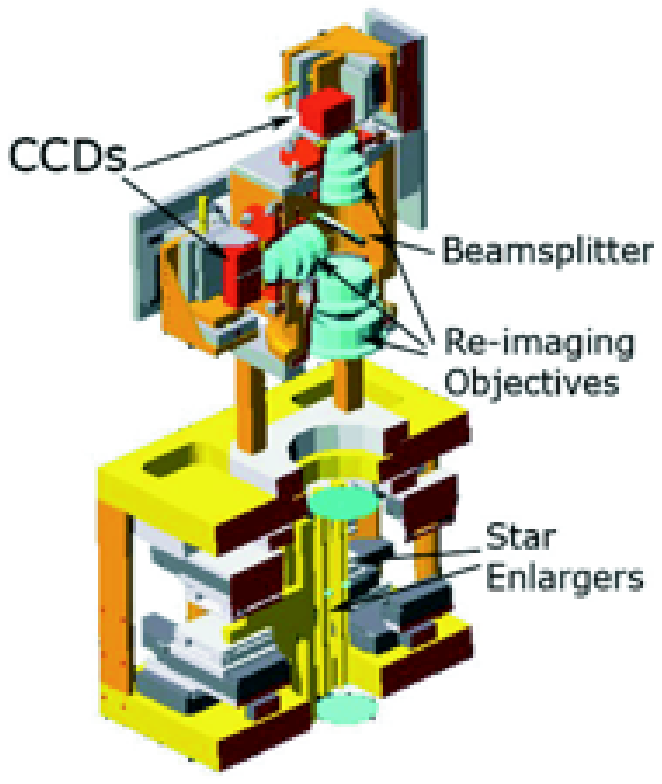}\label{fig:madcad}
		\includegraphics[height=2.5in]{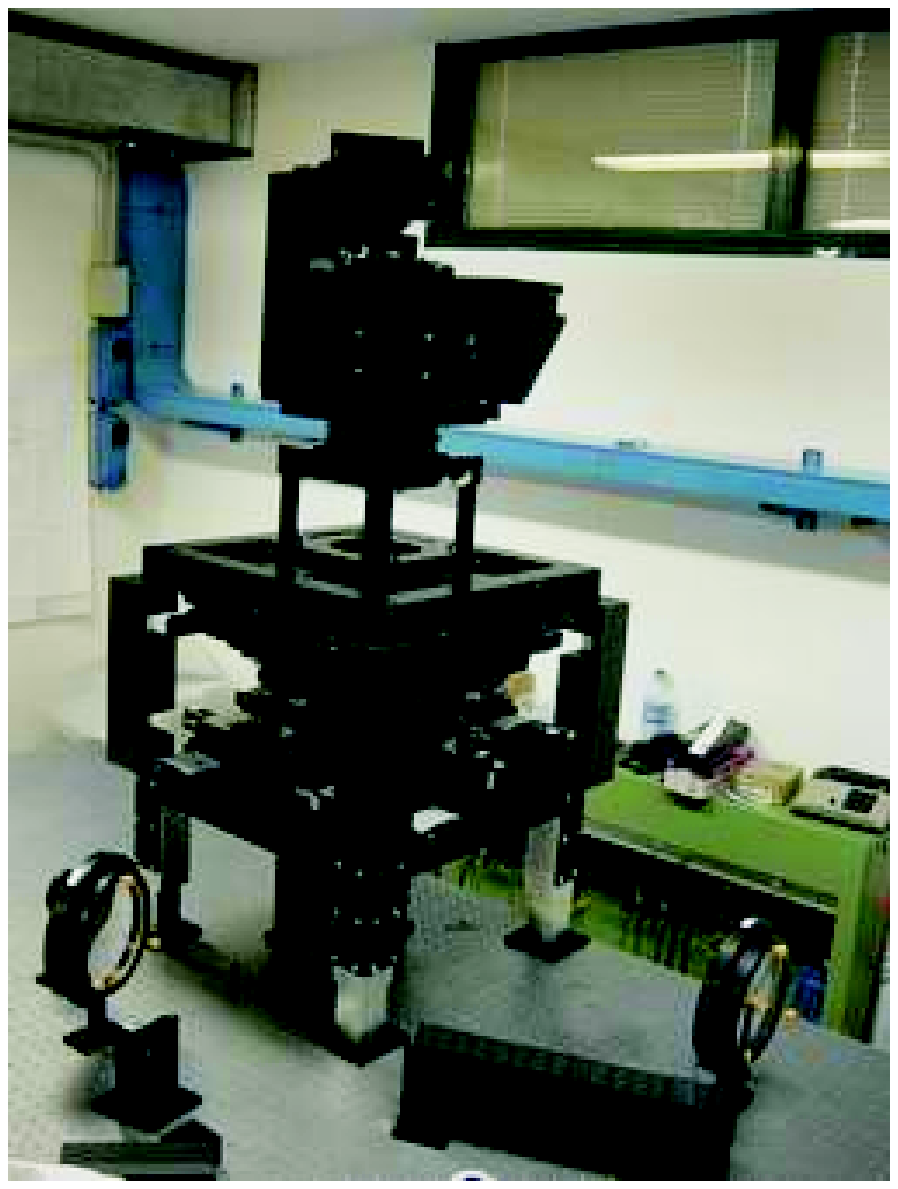}\label{fig:madphoto}
		\includegraphics[height=2.5in]{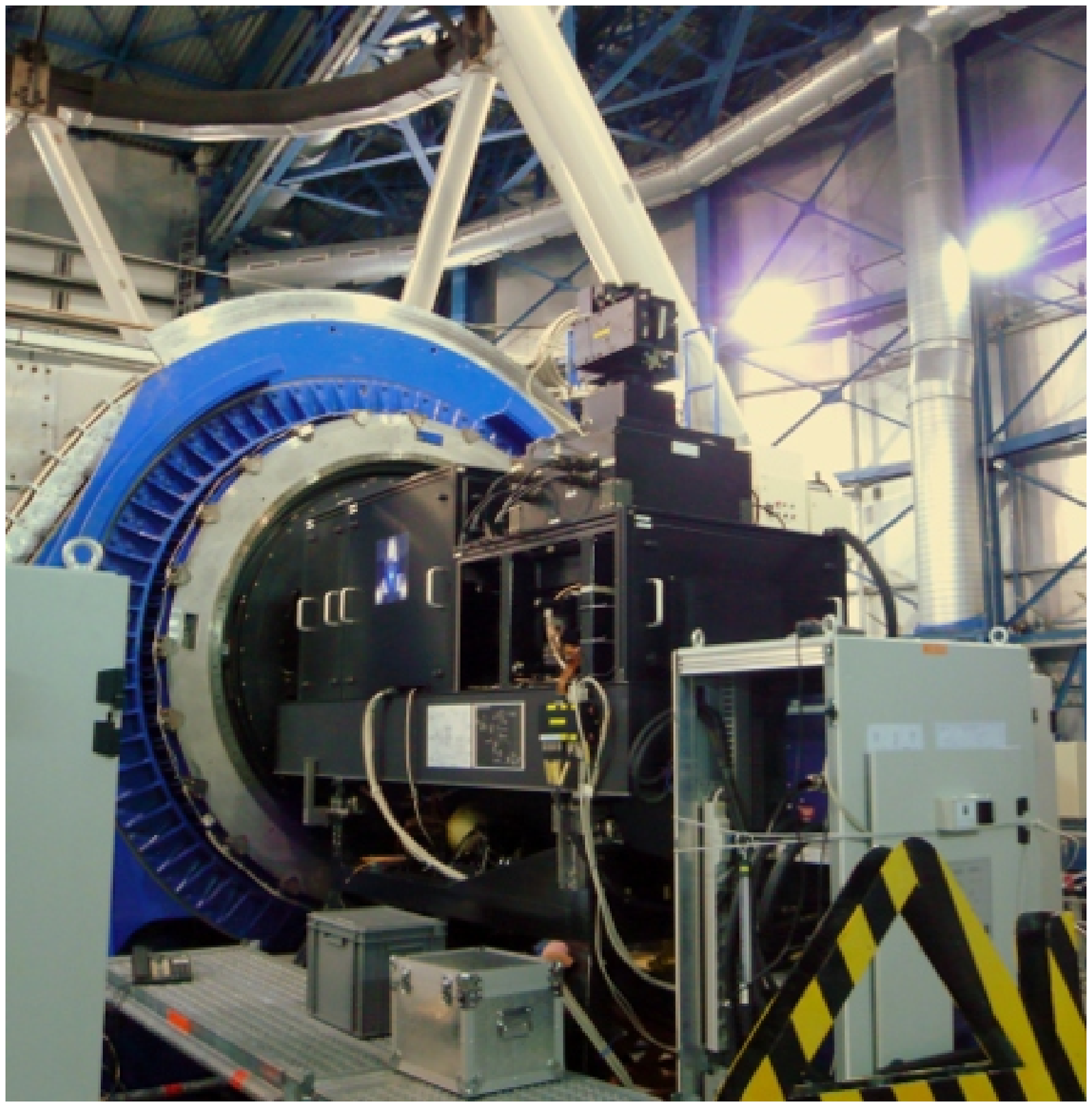}\label{fig:madon.eps}
   \caption[fig2] 
   { \label{fig:fig2} \label{fig:mad3d}
   These figures present a CAD
view and two real pictures
of the MAD Layer Oriented WFS.
   }
\end{figure}
In the LOWFS the light of each reference star is split into four beams by a pyramid placed on the focal plane and centered on the star. Then a reimaging objective projects the pupil images on a CCD (4 pupil images, one for each pyramid face). The different reference pupils are super--imposed according to the conjugation altitude and the stars directions mimicking the super-imposition of the pupils foot--print on the conjugated layer, realizing the optical conjugation of the sensor to the layer/deformable mirror.
In the LOWFS for MAD up to 8 pyramids can be positioned over the
2\arcmin FoV to catch the light from eight reference guide
stars, Figure~\ref{fig:mad3d}. 
In the LOWFS for MAD the reference stars pupils co--addition is optically performed on two different WFSs, which looks simultaneously to the four pupil images of all reference stars thanks to a 50/50 beam splitter placed on the pupil re--imager optical path. The pupil image corresponding to the F/20 WFS--input beam is too large to match the dimension of the wavefront sensing fast read out CCD, even if used with very fast re--imaging optics.
The two detectors were EEV39 with $80\times 80$ pixels, with 0.024mm pixel--size corresponding to a $1.92\times 1.92$mm sensing region:
pupil shrinking was necessary to match CCD size. The trick used is described in the cited paper\cite{tubetti} and it consists in enlarging the focal ratio only on the optical path of the reference stars in front of each pyramid. This goal was achieved using a co--moving optical train for each pyramid: two small diameter achromatic doublets have been used to obtain a new F/300 focal plane in correspondence of the pyramids vertex. The second lens diameter, or better its mounting, fixes the minimum
separation between two reference stars. Two adjacent references cannot stay closer than 20\arcsec (centre--centre) on sky.
The final pupil sampling on the LOWFS CCD were $8\times 8$ and $7\times 7$ respectively for the ground and high layer WFS, using different on chip pixel binning (2 and 4 respectively). In the MAD case a finer spatial sampling was not useful being the number of actuators on the two DMs 60 (on each) matching approximately the number of sub-aperture on the sensed pupil on ground and high layers CCDs using the binning above.

The two CCDs are positioned in focus with great accuracy ($\sim 1 \mu m$) by linear stages in order to properly conjugate the sensor to the deformable mirrors (and the corresponding layer altitudes, anyway the layer altitude is not critical). The motorized stages are identical to those used to move over the 2\arcmin FoV the mounting of the optical trains composed by the pyramid and by the 2 achromatic doublets: these 8 opto-mechanical systems are screwed to as many xy couples of linear stages.
\section{The observations}
In the commissioning plan the first three night were conceived to be merely technical  and devoted to evaluate the correction performance of the LOWFS under different system configurations and to bring the Layer Oriented system to a level of functionality sufficient for executing the GTO nights (6 nights from September 24th).  
But in the three technical nights we had only the first night with reasonably good seeing conditions (seeing in V band measured by the ESO-Paranal seeing monitor between 0.8 and 1.0 arcsec). Thanks to the efforts of the whole MAD team we succeeded to perform all the tests starting from Single Conjugate AO, passing trough Ground Layer AO and finally Multi-Conjugate AO. In a previous SPIE conference we already presented the results relative to the most relevant examples of the three cases\cite{2008SPIE.7015E.155A}.
Being the scope of this paper to offer an update of the results, we complete the already presented data sets with all the other relevant data taken in the other good seeing condition nights we had during the GTO time (the 27th September night).

But before starting this list we would like to recall the main problems occurred during the commissioning in order to properly justify the performance presented below. 

One known problem was the failure of one of the 8 optical trains (the star enlarger). The smallest of the two achromatic doublets unstuck and we used the other seven only.
Of the seven star enlargers two of them were not usable out of a restricted sky-projected region of about 10-15arcsec diameter close to the mid field at about 30arcsec from the centre of the field of view because the xy motorized stages on which they were mounted were introducing a large tip of the optical train, out of the specification (this tilt corresponding to a rigid shift of the four pupils on the CCD). At the very end we had the possibility to exploit the full 2\arcmin FoV to look for reference stars, but we had to find the proper combination of pyramids to be used, this required an extra work for the targets definition and references selection.

On the side of the MAD bench we noticed a source of light pollution inside the bench itself which unfortunately was generating a pupil image close to the entering F/20 focal plane. Anyway the contamination due to this light was not affecting the performance in a sensitive way. The major light pollution problem we had was on the side of IR channel: it was some un-filtered light which was passing (by reflections) into the dewar contaminating part of the imaged FoV, since this contamination was depending on the position of the camera on the field the sky subtraction from the scientific images was particularly difficult being impossible to eliminate it by simply making a median of the dithered images or considering sky frames taken pointing away the telescope. 

Anyway the most affecting problem was a software error in the control of the position of the optical derotator (mounted on the bench), which was resolved in a later commissioning session of the instrument. 
Unfortunately a rotation of the field is a mode not controllable by the DM and corresponds to a shift (different for each pyramid) of the reference stars PSF with respect to the pyramids position. In the laboratory test phase we already noticed the importance of a correct pyramids position on the focal plane: the rotation makes the pyramid working out of the linear regime, and, exceeding the 2 degrees was corresponding to have the stars out of the (small) 0.95arcsec FoV of each pyramid (for a pyramid at the center of the field this effect is invisible, 2deg refers to stars at 30arcsec from the centre). This problem was more evident at high elevations (larger than 70deg) at which rotator speeds are higher. The only solution we found to compensate for the, at that time unknown, field rotation was to recenter the pyramid positions by moving the xy linear stages. The re-centering of the pyramids was leading us to repeat the final part acquisition procedure, which includes the automatic centering of pyramids on the stars by minimizing the tip-tilt signal for each reference and in open loop. In some case, for large rotation errors, we needed to repeat the complete procedure looking for the stars on the technical camera used for 2\arcmin$\times$2\arcmin field imaging. Both procedures were unfortunately quite time-demanding being the linear stages used particularly slow (2mm/sec corresponding to about 1.5arcsec/sec).

\section{Multiconjugate adaptive optics observations}
A full characterization of the MCAO performance as we performed in laboratory before the delivery of the instrument to Paranal was not possible during on sky test. The main limitation was the bad seeing conditions encountered during LOWFS commissioning. The design of the LOWFS was made upon considerations based on median seeing condition of the Paranal site, according the measurements campaign varying between 0.65\arcsec to 0.7\arcsec, and on the achievable performance offered by the system, considering the number of actuators available and correction framerate. Nevertheless the last measurements on the median seeing pose this value around 0.70\arcsec\cite{2008Msngr.132...11S}, showing at the same time a discrepancy between DIMM measurements and the actual seeing at the focal plane of the UTs due to a strong ground layer in the first 50meters unfortunately we had good seeing conditions (better than 0.8\arcsec)only in two of the nine assigned nights.
A note on the way we acquired IR images: the FoV imaged by the CAMCAO camera was a square of 57\arcsec$\times$57\arcsec, therefore to cover the full 2\arcmin corrected FoV we needed to move the CAMCAO over the field using the couple of linear stages it was mounted on. To acquire the full field we took 4 frames on a square geometry and one on the on axis direction.

Later we present the four useful cases in MCAO while for the those observed in bad seeing conditions we just say we achieved a factor $2.5 \pm 0.5$ Full Width at Half Maximum (FWHM) shrinking in Ks: just to make an example in the case of the globular cluster M55 the initial Brakett-$\gamma$ band open loop image showed 0.75\arcsec FWHM and closing the MCAO loop correcting 50 modes (the total of Ground and High) and 6 stars we achieved 0.24\arcsec average FWHM (the magnitudes of the stars in V band were 12.7, 13.8, 13.6, 13.3, 13.7 and 14.2). Notice that this $2.166\mu m$ wavelength seeing value scaled to the V band is 1.00\arcsec and well fits the ESO DIMM value registered for the time of that specific observation corresponding to 1.10\arcsec. This seeing value is marginally out of the WFS working range, in fact the $\sim$ 1.\arcsec is just 0.1\arcsec larger than the pyramids FoV. 
  \begin{figure}
   \begin{center}
   \begin{tabular}{c}
   	\includegraphics[height=3.5in]{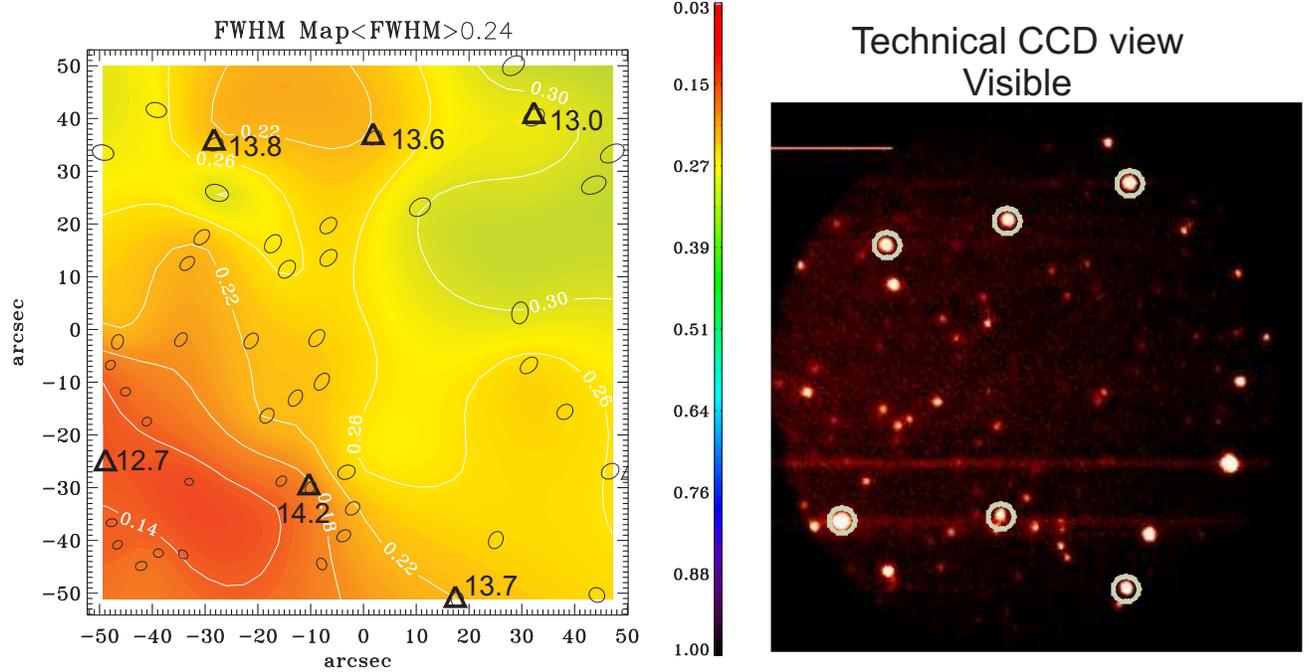}
   \end{tabular}
   \end{center}
   \caption[fig3]
   { \label{fig:fig3} \label{fig:m55}
   In these figures a representative case of how the LOFWS performed in bad seeing conditions. We observed the Globular Cluster M55. The measured Br$\gamma$ band seeing was 0.75\arcsec. On the left the FWHM map of the $2 \times 2$ \arcmin FoV. On the right the visible image taken by the technical CCD use during the reference acquisition procedure. It's interesting to notice that the combined effect of the de-rotation error in the image is visible. In fact in Layer Oriented the applied correction is weighted by the brightness of the reference stars: if a rotation error is present the focal plane projection of the field tends to be more centered (with respect to the pyramids) on the region where the brightest stars are. In this case the brightest source is on the bottom left, exactly where the best performance are, while at 180degrees we found the poorest. The most probable explanation is that the reference stars on the top right were decentered with respect to the corresponding pyramids generating an extra source of noise.
   }
\end{figure}
\subsection{Three bright stars}
At the end of the first technical night (the 21$^{\rm st}$) we succeeded to close the MCAO loop:
we selected as references three bright stars  well spaced (magnitudes 11.059, 11.157 and 12.072) in the \bytwo FoV  and we performed the acquisition procedure, pointing the telescope in RA 03:53:25.86 dec -50:07:20.9.
The original seeing value was 0.46\arcsec$\pm$0.03\arcsec measured on the CAMCAO camera, to be compared to the 1.45\arcsec (in V band) measured by the seeing monitor (to be noticed that $\lambda^{1/5}$ seeing scaling gives 1.08\arcsec scaling from the wavelength $2.166\mu m$ to V band, such a big difference stays, probably, on a strong ground layer below the UTs but above the DIMM).
\begin{table}[h]
\caption{The following table summarizes the results obtained in GLAO and MCAO mode on the first bright MCAO case. We measured Full width at half Maximum, FWHM, by fitting Moffat Function on the ${{\rm Br}}\gamma$ images.  The Ensquared Energy in 0.1\arcsec (EE$_{0.1}$) is listed showing a gain $\sim \times 3$ and $\sim \times 4.5$ with respect open loop case respectively for the GLAO and MCAO. SR$_{{\rm Br}}\gamma$ and $\sigma_{{\rm{V,DIMM}}}$ are respectively the seeing FWHM measured by the DIMM during the exposure (in V-band) and the measured Strehl Ratio. The field is a mosaic of 5 exposures (30~seconds each) in order to cover the \bytwo.}
\label{tab:MCAO0}
\begin{center}
\begin{tabular}{|l|r|r|r|r|r|r|r|} %% this creates two columns
%% |l|l| to left justify each column entry
%% |c|c| to center each column entry
%% use of \rule[]{}{} below opens up each row
\hline
\rule[-1ex]{0pt}{3.5ex}                           & M$_{V}$              & FWHM [\arcsec]   ${{\rm Br}}\gamma$       & EE$_{0.1} [\%]$ ${{\rm Br}}\gamma$ &SR$_{{\rm Br}}\gamma$ [\%] & $\sigma_{{\rm{V,DIMM}}}$ [\arcsec] @V\\
\hline
\rule[-1ex]{0pt}{3.5ex} Open Loop               &  -                     & 0.45           &  4.7         &    1.8             &1.48\arcsec\\
\hline
\rule[-1ex]{0pt}{3.5ex} GLAO Closed Loop & 11.059, 11.157, 12.072 & 0.17$\pm$0.02  & 14.9$\pm$2.1 &   9.2$\pm$2.6     & 1.46\arcsec\\
\hline
\rule[-1ex]{0pt}{3.5ex} MCAO Closed Loop & 11.059, 11.157, 12.072 & 0.12$\pm$0.04  & 23.3$\pm$3.9 &  17.3$\pm$9.1     & 1.39\arcsec\\
\hline
\end{tabular}
\end{center}
\end{table}

\begin{figure}
\centerline{\includegraphics[height=3in]{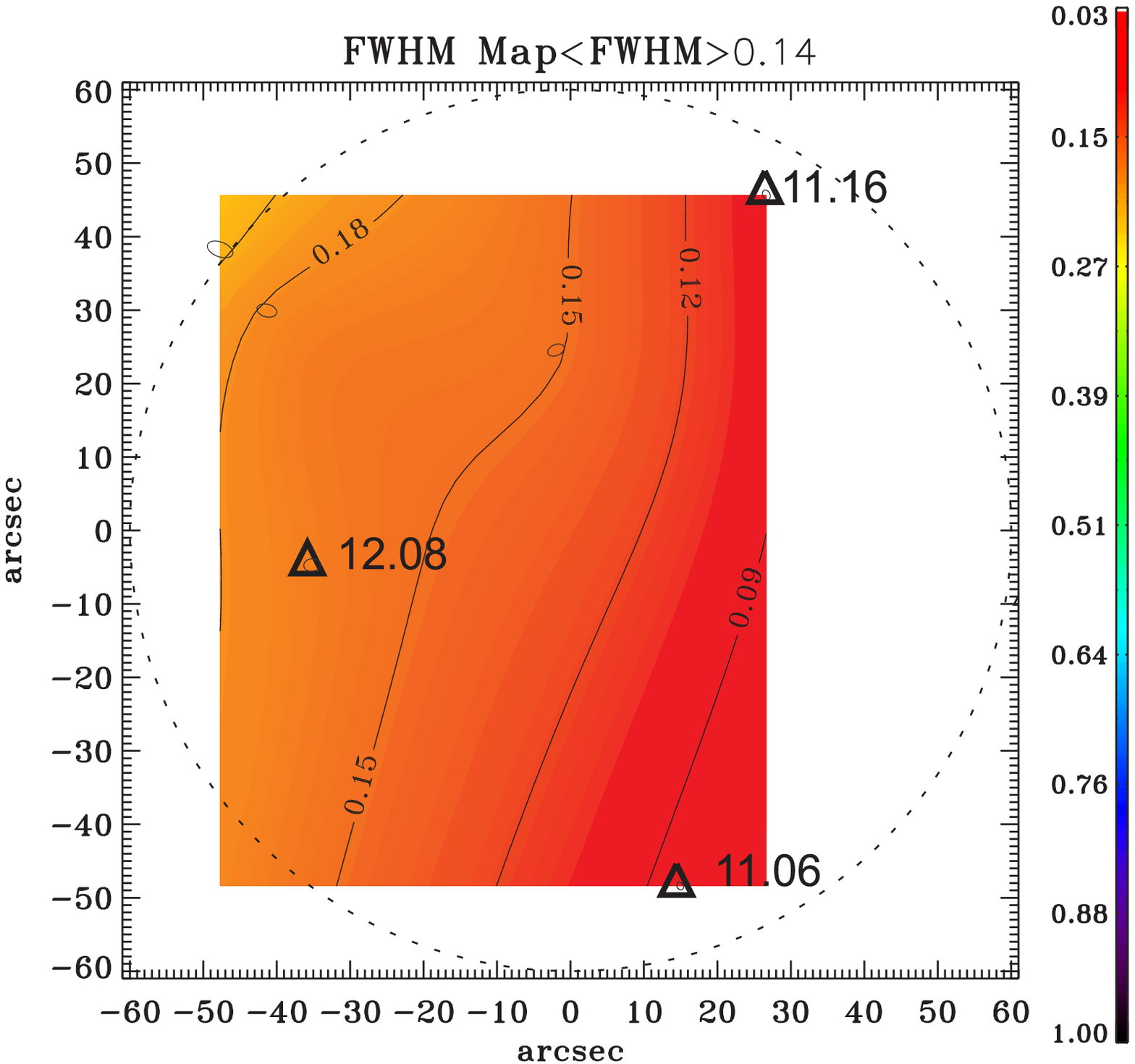}
\includegraphics[height=3in]{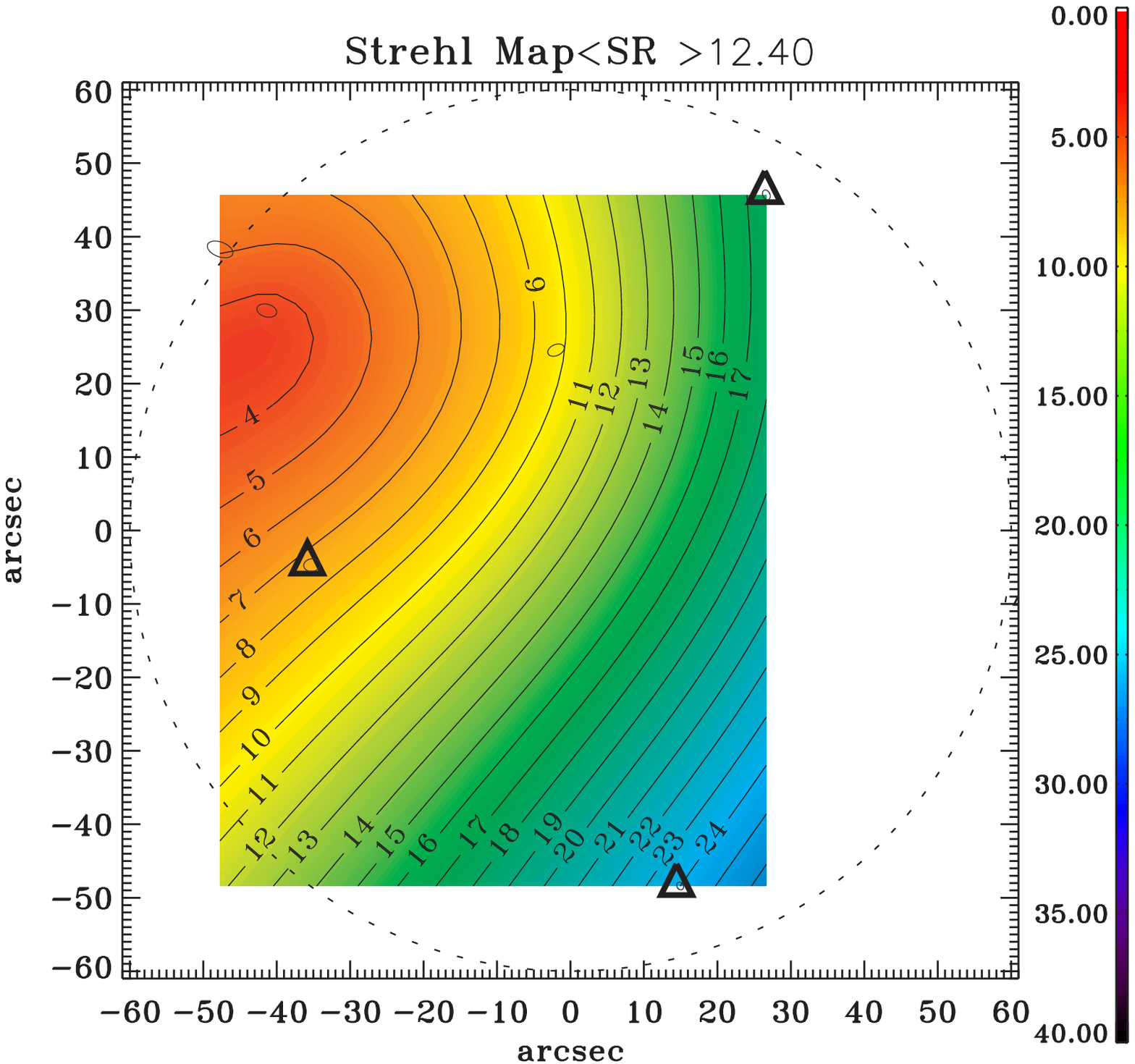}}\caption{\footnotesize
This pictures represent the analysis on the first (and not optimized) MCAO close loop made in Layer Oriented mode. On the left picture the positions of the reference stars (triangles) with respect to the observed FoV are superimposed to the measured FWHM map. Close to the reference stars is written the corresponding V band magnitudes. The ellipses represents the size (enlarged) and orientations of the PSF such as have been fitted using Moffat Functions. On the right an image of the Strehl Ratio map. The initial seeing was 0.45\arcsec measured on the open loop images.}\label{fig:ast1}
\end{figure}
\begin{figure}
\centerline{\includegraphics[height=3in]{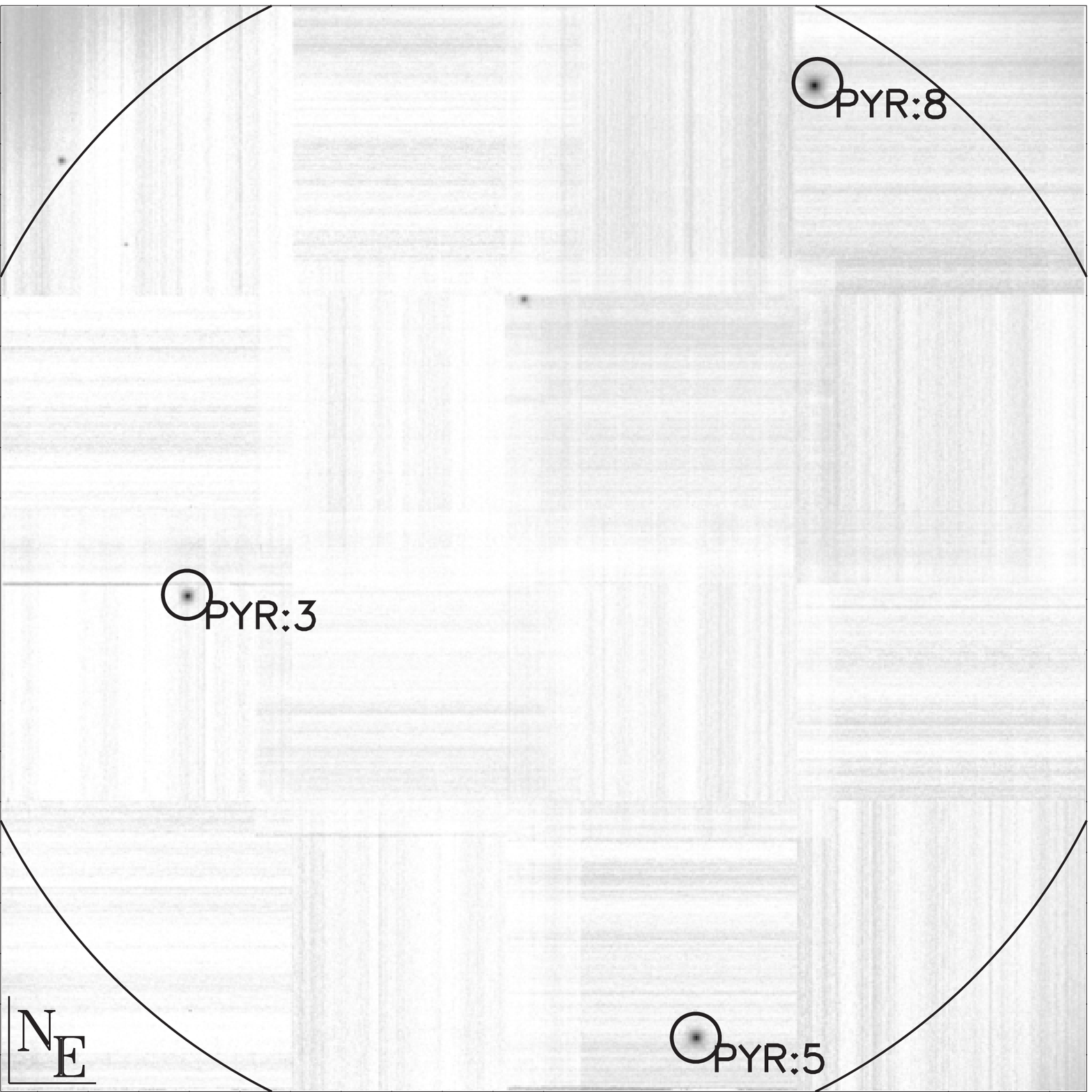}\includegraphics[height=3in]{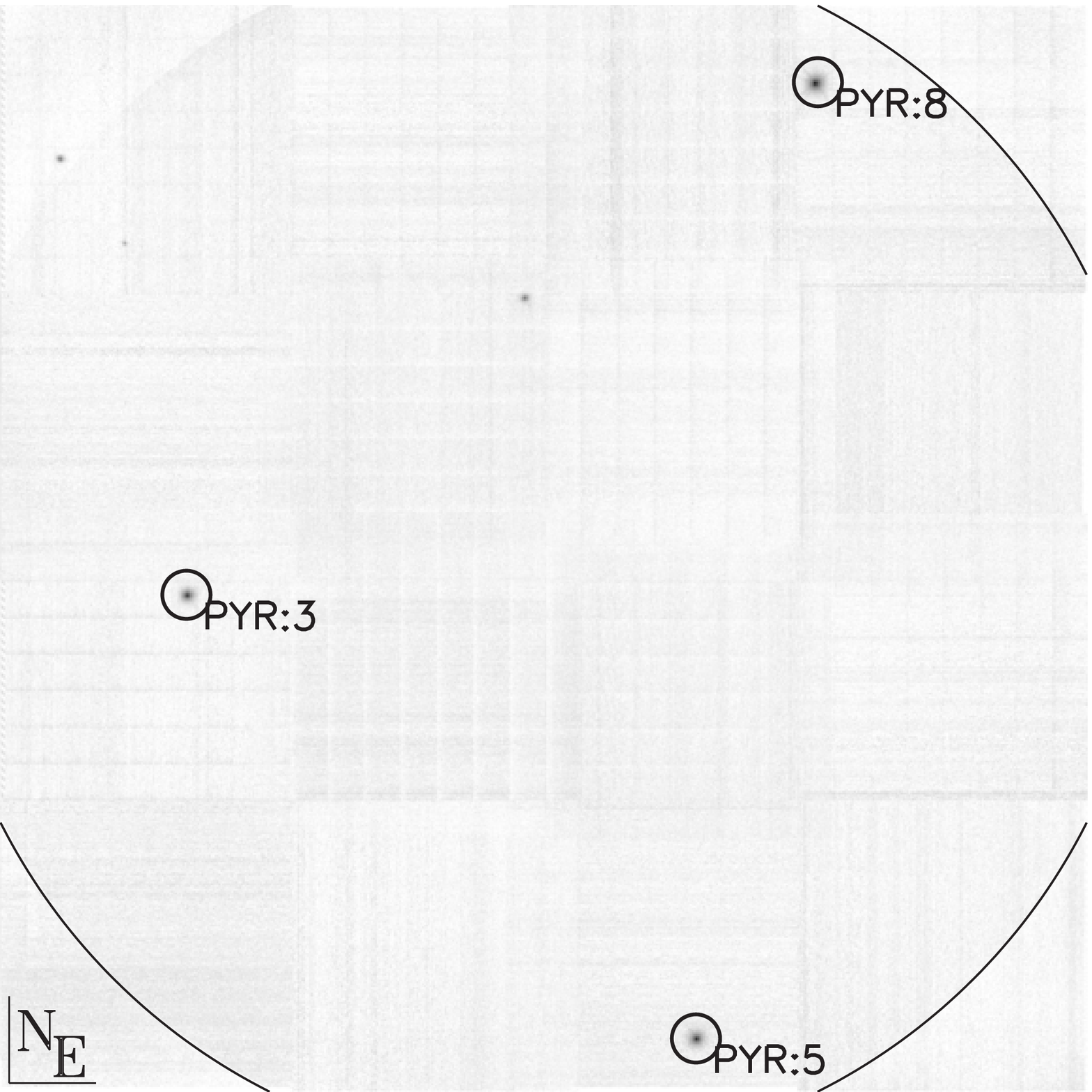} }\caption{\footnotesize
The picture shows the IR mosaics of the \bytwo field relative to the first (and not optimized) MCAO close loop made in Layer Oriented mode. On the image the circles identify the 3 reference stars, with the corresponding pyramid identification numbers. The left image shows the GLAO mosaic the right the MCAO, comparing the two we notice the error in derotator position. See text.}\label{fig:ast1im}
\end{figure}
 
Unfortunately also in this case we found an error in the position of the optical derotator: comparing the mosaic of 5 images (30seconds each) composing the GLAO case and MCAO case  (see Figure~\ref{fig:ast1im}), taken at about 6~minutes of time difference, shows a shift in the position of the stars. This shift is compatible with a rotation of the field of the order of 0.2degrees.
This error explains the Strehl Ratio (SR) value drop on the last frames taken, corresponding to the faintest reference star (see Figure~\ref{fig:ast1}), which shows an anomalous SR difference with respect to the brightest one. While the brightest was well centered the other two were slightly off-positioned with respect to the corresponding pyramids. To be noticed that being the end of the night there was not time available for loop gains optimizations.
\subsection{NGC6388}
For the globular cluster NGC6388 we did two different pointings: one on the globular cluster center the 26$^{\rm th}$ September (RA~17:36:17.86, dec~-44:44:05.60) and the others on a external region of the same cluster, the 27$^{\rm th}$, 28$^{\rm th}$ and 29$^{\rm th}$ September (RA~17:36:22.86, dec~-44:45:35.53). Unfortunately only the night of the 27$^{\rm th}$ had acceptable seeing conditions: the results of this night will be discussed in the following while for the others we just say that we obtained only a FWHM improvement and Ensquared Energy concentration gain with respect to open loop seeing PSF, however far in absolute terms from the results obtained in the best night.
The scientific results of the analysis of the 31 images taken with DIT 10 seconds, and NDIT 24 (20 frames) and NDIT 12 (11 frames) are presented in the cited papers\cite{2009A&A...493..539M,2009MmSAI..80..139M}. In the analysis reported here we excluded the last 5 frames, which present much lower performance in terms of SR (6 to 8 percent with respect to an average of 14 percent). Also in this case we discovered that these frames were slightly rotated with respect to the others.

We estimated the open loop seeing both using DIMM data retrieving 0.55\arcsec (in V) and directly measuring the FWHM of the stars in the ``sky'' frames used for data reduction corresponding to 0.33\arcsec (this time in Ks). 
\begin{figure}
\centerline{\includegraphics[height=3in]{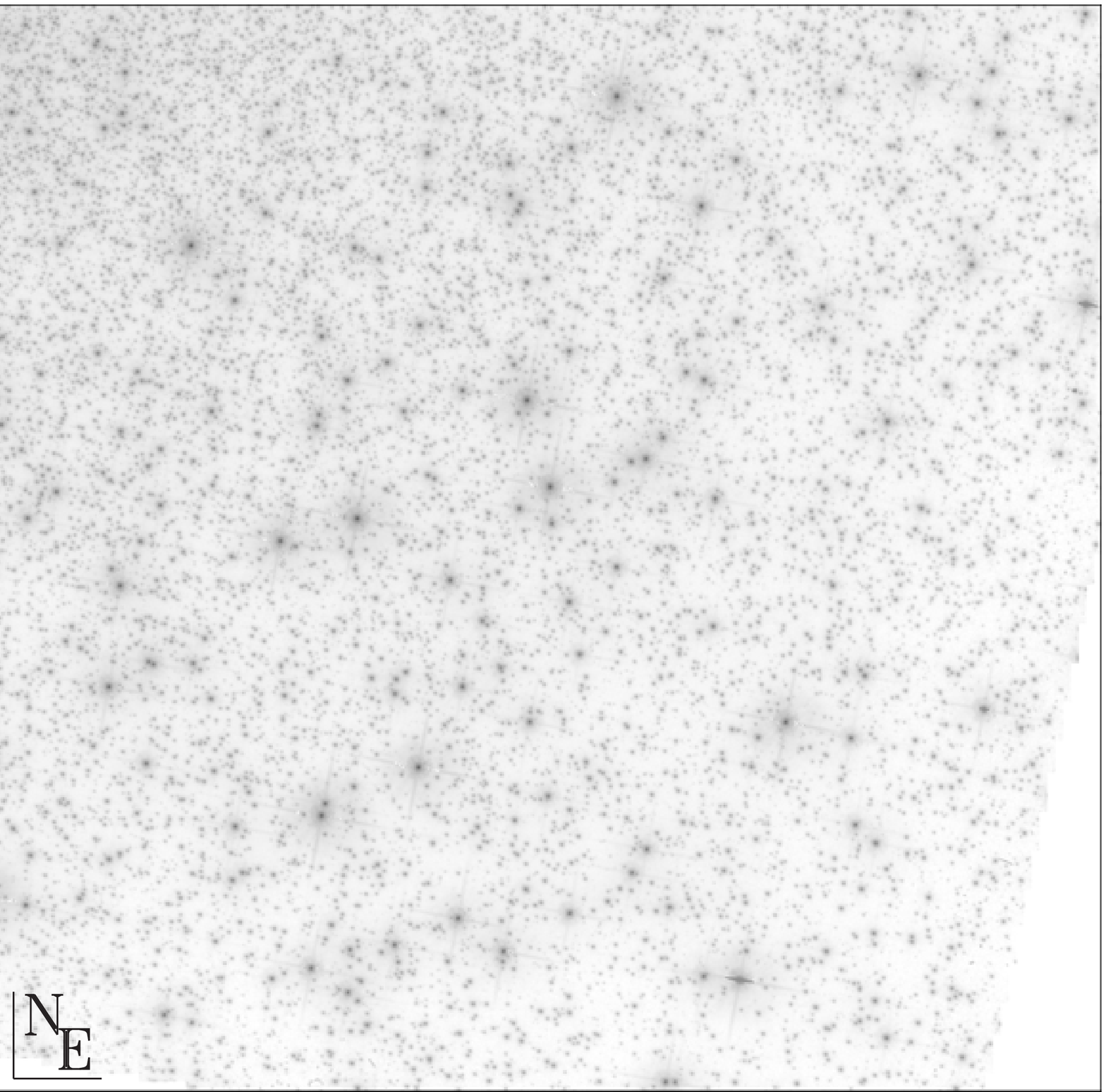}
\includegraphics[height=3in]{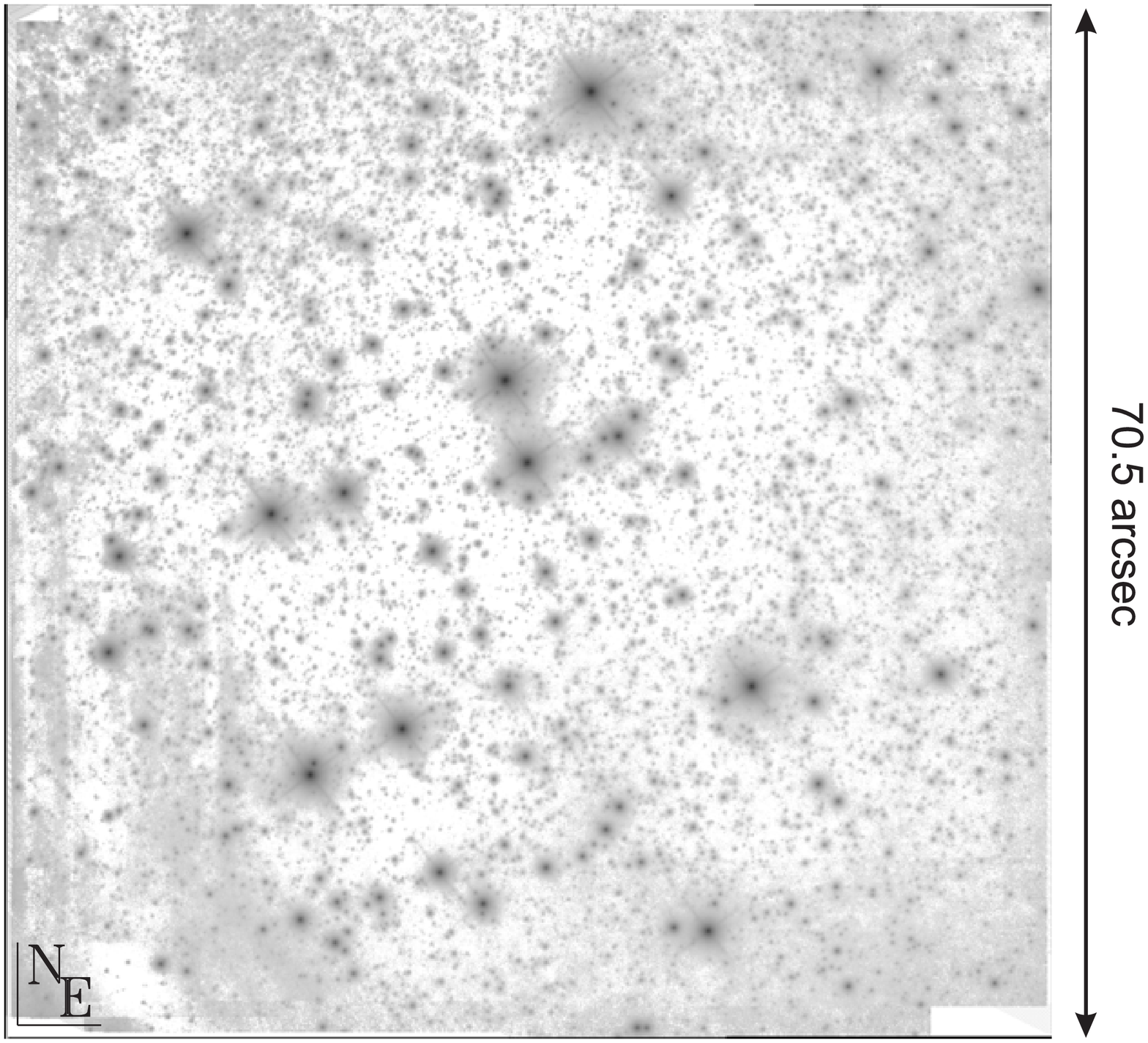}}\caption{\footnotesize
Here a visual comparison between images of the field composed using HST F606 filter images (left) and LOWFS-MAD Ks (right). The two cuts have been rotated and flipped in order to show the same part of the external region of the globular cluster NGC6388.}\label{fig:6388a}
\end{figure}

\begin{table}[h]
\caption{The following table summarizes the results obtained in MCAO mode on the external region of the globular cluster NGC6388. FWHM are measured by fitting Moffat Function.  SR$_{{\rm Ks}}$ and $\sigma_{{\rm{V,DIMM}}}$ are respectively the measured Strehl Ratio in SR ${{\rm Ks}}$ and the seeing FWHM measured by the DIMM during the exposure in V filter. In this case the visual magnitudes M$_{V}$ are Hubble Space Telescope F606W-filter photometry data. The presented values have been estimated on the composite image made by summing the 26 Ks-frames (the ones less affected by the optical derotator error) for an overall exposure time of 5520 seconds. $\sigma_{{\rm{Ks,CAMCAO}}}$ is the seeing FWHM measured on open loop images available (the sky frames).}
\label{tab:MCAO1}
\begin{center}
\begin{tabular}{|l|c|c|c|c|c|c|c|c|} %% this creates two columns
%% |l|l| to left justify each column entry
%% |c|c| to center each column entry
%% use of \rule[]{}{} below opens up each row
\hline
\rule[-1ex]{0pt}{3.5ex}  & M$_{V}$              & FWHM [\arcsec] & EE$_{0.1''} [\%]$ &SR$_{{\rm Ks}}$ [\%] & $\sigma_{{\rm{V,DIMM}}}$ [\arcsec] & $\sigma_{{\rm{Ks,CAMCAO}}}$ [\arcsec]\\
 &                & @Ks & @Ks & @Ks & @V& @Ks\\
\hline
\rule[-1ex]{0pt}{3.5ex} NGC6388 & 15, 15, 15.6, 15.7, 16.2 & 0.13$\pm$0.01  & 21.2$\pm$5.2 &   14.1$\pm$2.5     &  0.55 & 0.33\\
\hline
\end{tabular}
\end{center}
\end{table}

\begin{figure}
\centerline{\includegraphics[height=3in]{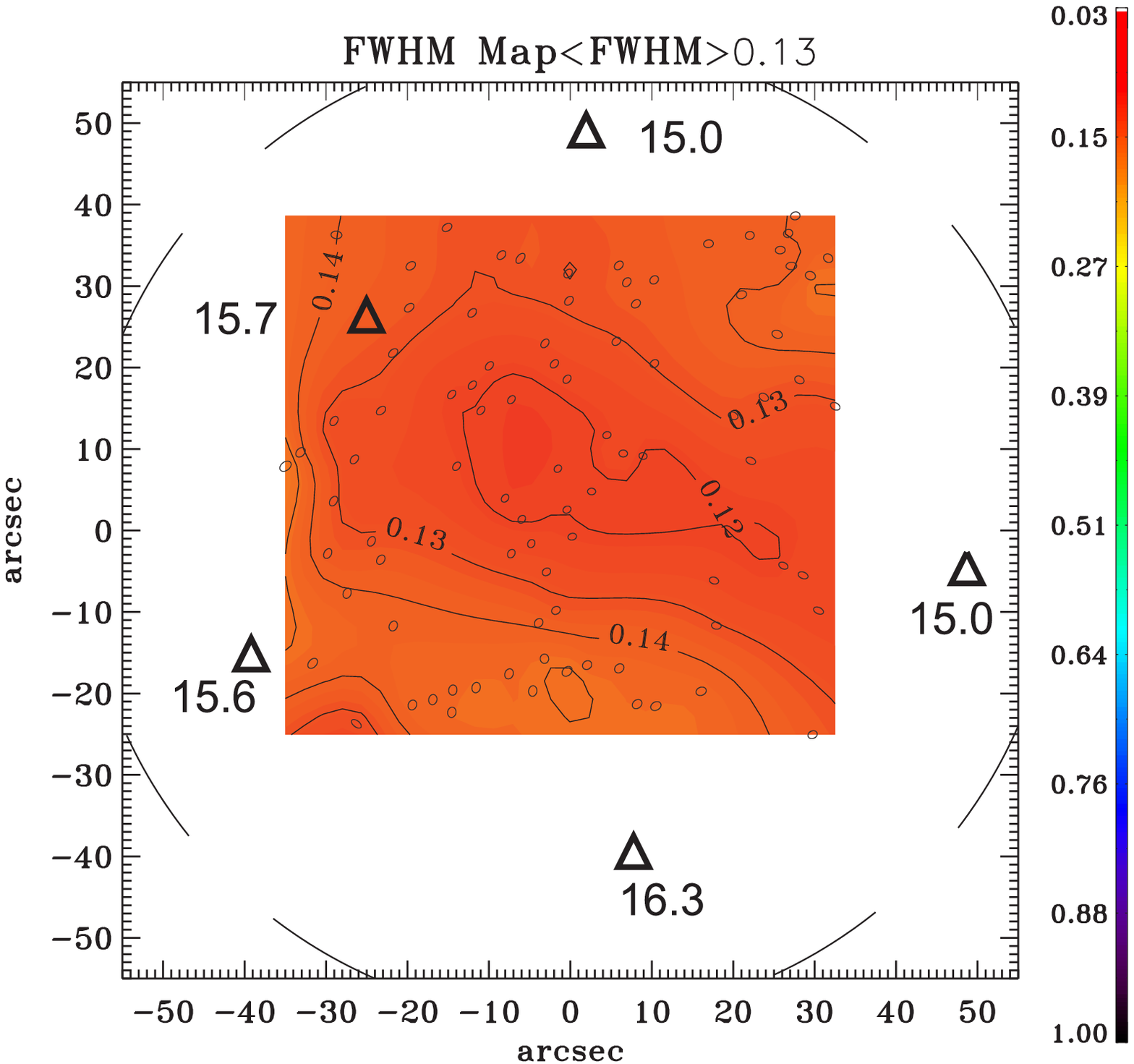}
\includegraphics[height=3in]{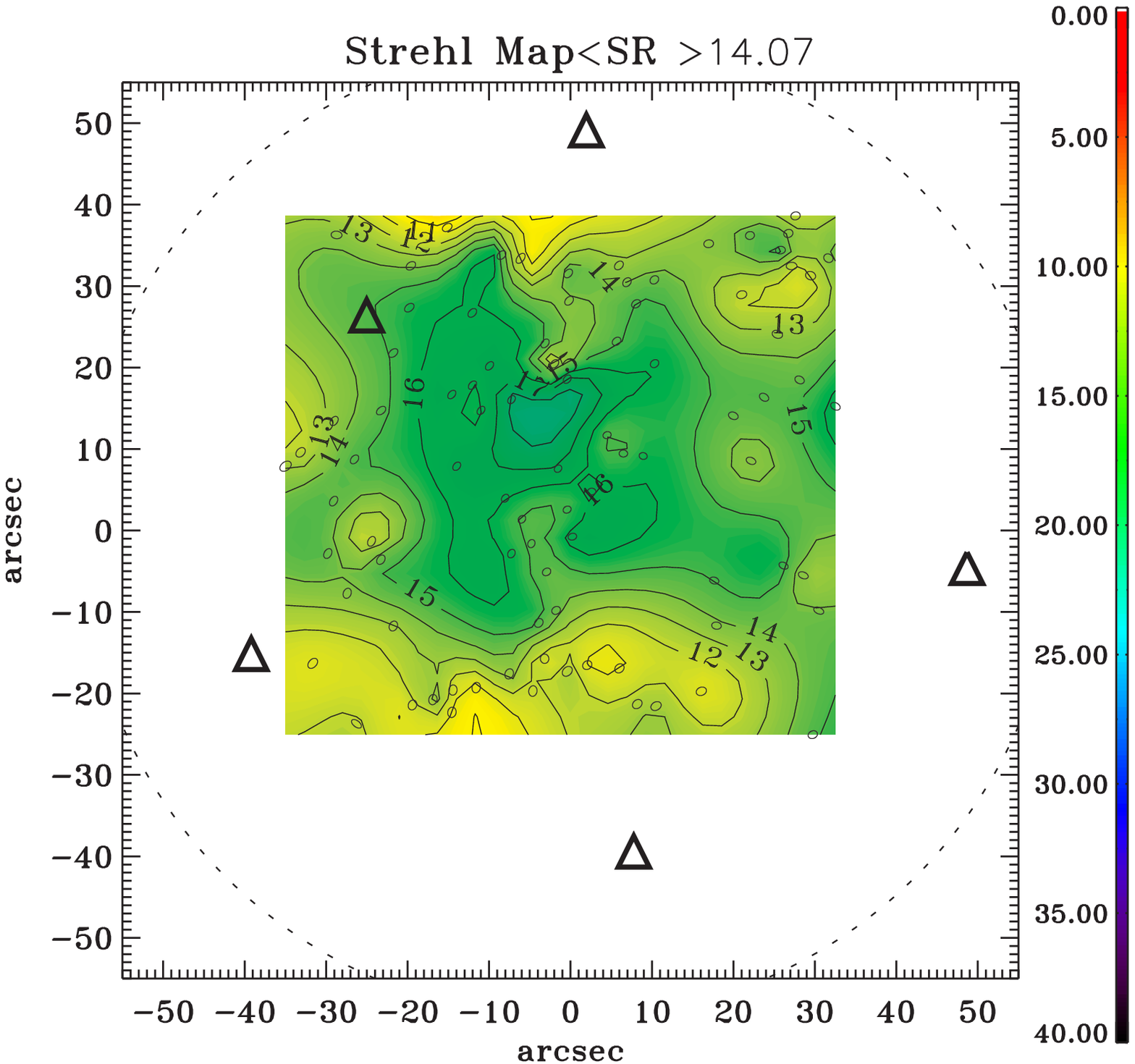}}\caption{\footnotesize
On the left picture the positions of the reference stars (triangles) are superimposed to the measured FWHM map for the observations relative to the cluster NGC6388. The corresponding HST-F606W filter magnitude is written close to the reference stars. On this region the FWHM are close to the diffraction limit, nevertheless the faint magnitudes of the reference stars, which integrated magnitude is 13.67. The ellipses represent the dimensions (enlarged) and orientations of the PSFs such as they have been fitted using Moffat Functions. On the right a map of the measured SR (Ks).}\label{fig:6388b}
\end{figure}

\subsection{PKS 0521-365}
We observed a BL Lac object, the radio source PKS 0521-365 (RA~05:22:57.985 and Dec~-36:27:30.85) in MCAO mode using the LOWFS\cite{2009A&A...501..907F}. The Layer Oriented system used three different reference stars to perform wavefront sensing, which present respectively R-band magnitudes 13.6, 13.0 and 11.4 (magnitudes taken from the USNO-B catalog), corresponding to an overall integrated magnitude of 11.1. MAD wavefront sensing CCDs sensitivity is centered in the V band, while the above magnitude are in R: considering B magnitude from the same catalog we can estimate an overall 11.9 V integrated magnitude. 
Actually the brightest reference is the core of the BL Lac, which can be considered as a star-like source for wavefront sensing. The telescope was pointing close to the Zenith, with airmass 1.05, and the ESO DIMM seeing monitor measured an average 1.1 $\pm 0.1$ arcsec seeing in Vband (rescaled to the Zenith). 

\begin{figure}
\centerline{
\includegraphics[height=3in]{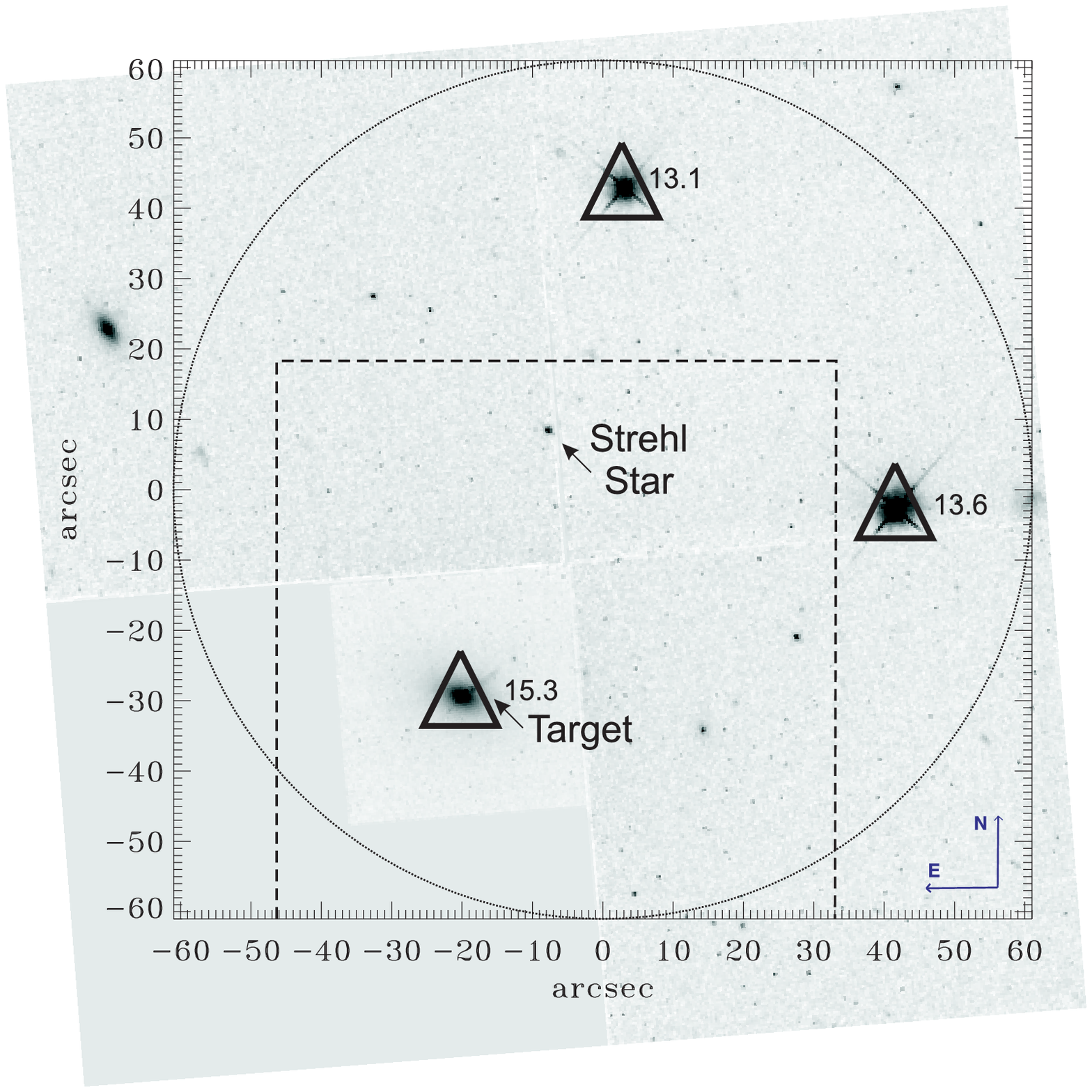}\includegraphics[height=3in]{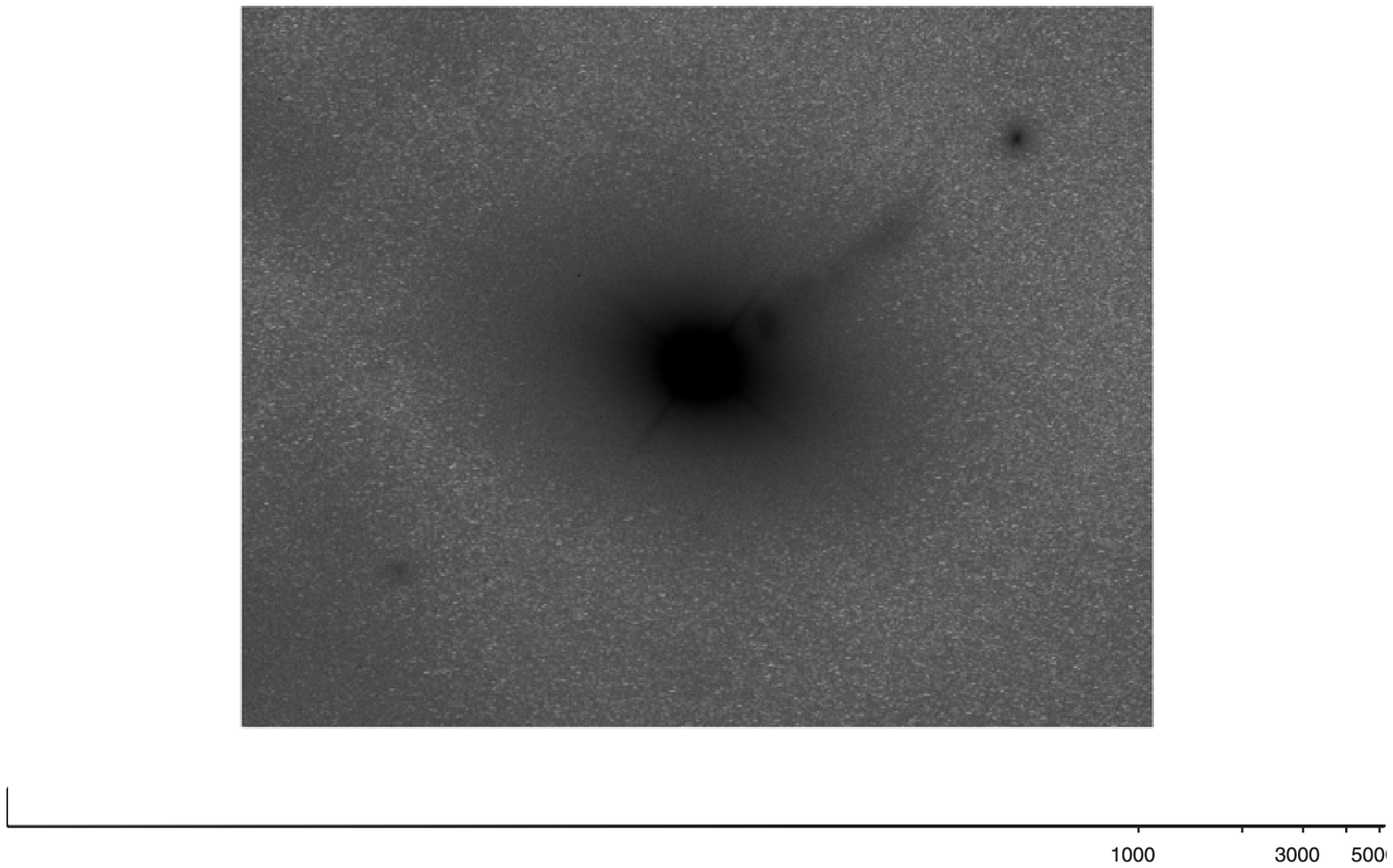}}\caption{\footnotesize
The image on the left shows the 2\arcmin field of view (circle) corrected by the MAD LOWFS system of the object PKS 0521-365. The area inside the dashed box was imaged by the NIR camera. The grey-scale image in background is a mosaic of HST + WFPC2 images (F702w filter). The triangles correspond to the LO AO reference stars. R-magnitude is reported for each star. At the centre of the field the arrow identifies the star used to compute the SR. On the right a log-scale subset of the IR image showing the details of the BL Lac object.}\label{fig:BL4}
\end{figure}

We computed the Strehl Ratio (SR) presented here, using a relatively bright star visible at about the center of the field (see Figure\ref{fig:BL4}): SR value cannot be computed on the core of the BL Lac because of the effect of the extended halo corresponding to the galaxy.

Images of the target were obtained following a 5-positions jitter pattern, with offsets of 5 arcsec, to ensure an adequate subtraction of the sky background. The entire data set consists of 20 exposures with slightly different positions of the pattern, for a total exposure time of 3600sec.
Dividing the exposure by four (as the four patterns repeated) we measure on the combined images this SR values: 14.4\%, 26.3\% 22.0\% and 15.9\%, all in Ks. The FWHM was between 90mas and 120mas.

\subsection{Deep Field}
\begin{figure}
\centerline{
\includegraphics[height=3in]{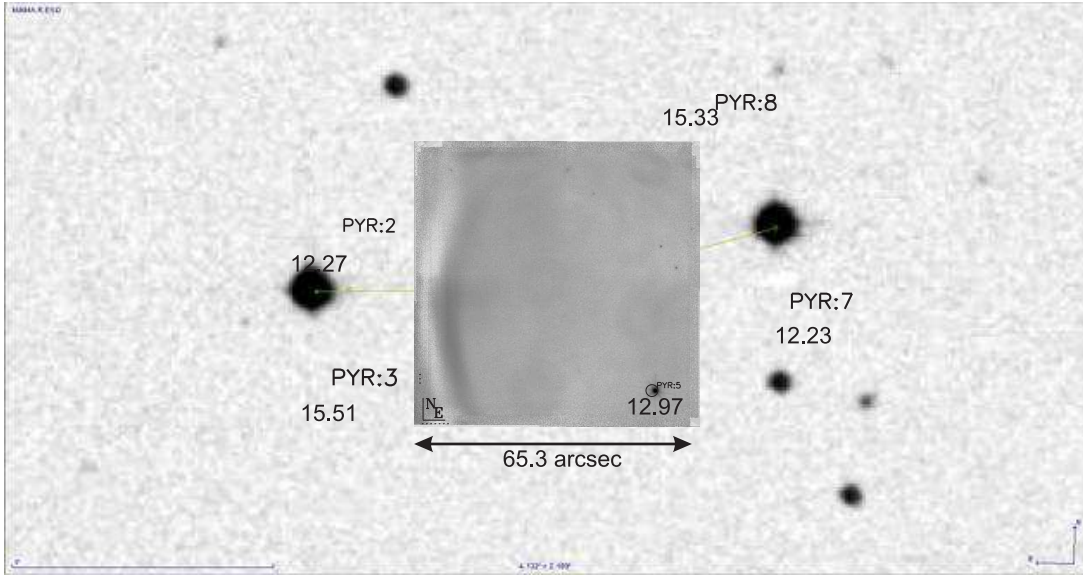}}\caption{\footnotesize
The image shows the 2\arcmin field of view corrected by the MAD LOWFS system as visible in the R band of the region observed for the deep field attempt. The area inside the box is the portion imaged by the NIR camera. The inset is the mosaic of all images available in Ks. The positions of the reference stars are defined by the pyramids identification names. R-magnitude is reported for each star found on the NOMAD catalog or the GSC. In the inset the circle identifies the star used to compute the SR (which is also a reference star).}\label{fig:DF1}
\end{figure}

We made an attempt of deep field imaging using a constellation of 5 reference stars pointing the telescope at 23:38:19.03 -30:59:03.2. We observed both in J and Ks band. In this case the five stars had magnitudes 12.23, 12.27, 12.97, 15.33 and 15.51, see Figure~\ref{fig:DF1}. Regarding the last and faintest star we had problems for the centering of the pyramid because of the poor signal we had at the slowest frequency we could read the WFS CCD (50Hz).
We observed using the same jitter pattern used for PKS 0521-365 (see the sub-section above). 
We started observation using GLAO mode passing to MCAO on the second jitter pattern realization. In this case we performed 6 jitters patterns, corresponding to 30 images, DIT 10 seconds and NDIT 6 for a total exposure time of 1800sec in Ks and as many in J. In Ks we measured SR between 7.4\% and 15.4\% measured on one of the reference star, the only one visible on the imaged field, with a 12.9\% mean value. In the best case we measured a FWHM of 94mas and an Ensquared Energy in 100mas of 29.5\%. 
We have not open loop data, but the ESO DIMM was measuring about 1\arcsec at the Zenith, while this observation was performed at $\sim$40degrees of telescope altitude (corresponding to about 1.5airmass).

\section{A comparison with laboratory data}
\begin{figure}
\centerline{
\includegraphics[height=4in]{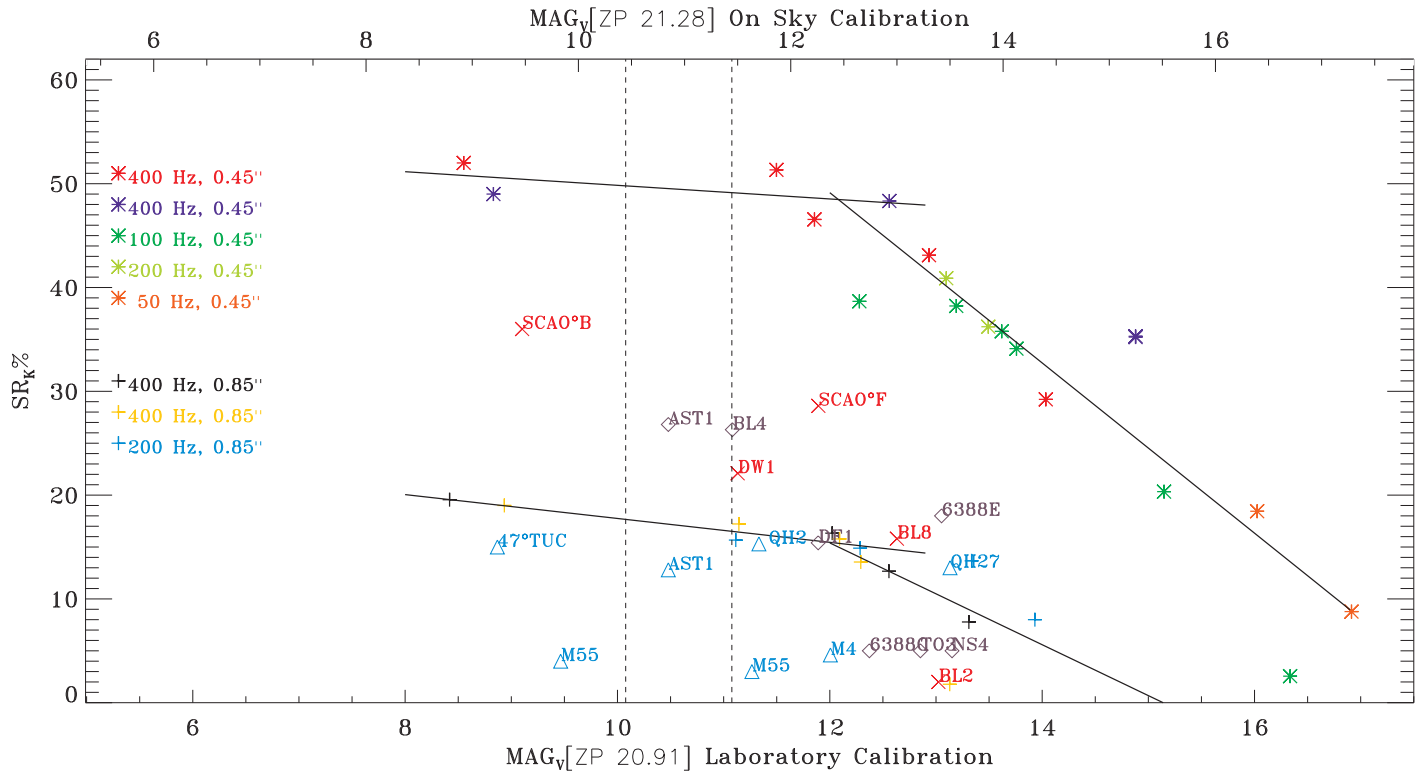}}\caption{\footnotesize
The image shows the data available for single pyramid limit magnitude determination (asterisk ``*'' and plus ``+'' signs, respectively corresponding to 0.45\arcsec and 0.85\arcsec seeing atmospheres (V band). The two solid curves are the best linear fits for the bright and faint magnitudes range. On these data are super-imposed the sky data, crosses ``$\times$'' for SCAO data, triangles for GLAO data and rhombuses for the MCAO. To be noticed that to obtain fair SCAO laboratory data should be considered the 50/50 beam effect here not taken into account. The two vertical lines represent the limiting magnitude achievable with multi SH, considering 3stars of 12th or 3 of 13th.}\label{fig:all}
\end{figure}

We performed extensive test to determine the limit magnitude achievable by the pyramid WFS. In this case we would like to place our (a few) on sky data to those already obtained in laboratory, see Figure~\ref{fig:all}. We see that the on sky on data (whatever mode SCAO, GLAO or MCAO) stay in the area between the two seeings data sets. This result was expected being the seeing we encountered on the good conditions nights lying in between those two values. Also in terms of limiting magnitude we were not so far from the limits found in the lab tests.
Comparing the fluxes and the relative magnitudes we found a difference of about +0.5magnitudes in favour of laboratory magnitude estimation (that means we had on sky fainter fluxes than expected).
\section{Conclusion}
In this paper we offered an overview of the most interesting MCAO cases we realized during both technical and GTO nights. We pointed out how the effect of the error in the optical derotator control software affected the observation both in terms of time and Strehl Ratio. The efficiency of the correction using faint stars has been proven and it represents a big issue thinking in terms of sky coverage: the data relative to the globular cluster NGC6388, an overall integrated magnitude 13.7 realized with 5 references fainter than 15th, represents a very important starting point for the future MCAO systems.
Various scientific papers\cite{Gullieuszik08,mignani08,2009A&A...501..907F,2009A&A...493..539M} have been published using the collected data and, the most important point, with MAD we had the opportunity to prove on sky the Layer Oriented MCAO technique and this was a team success. 
%%%%%%%%%%%%%%%%%%%%%%%%%%%%%%%%%%%%%%%%%%%%%%%%%%%%%%%%%%%%%
\acknowledgments     %>>>> equivalent to \section*{ACKNOWLEDGMENTS}       
Thanks are due to all ESO staff and collaborators, E. Marchetti, J. Kolb, R. Donaldson, E. Fedrigo, S. Tordo and R. Gilmozzi among the others.

%%%%%%%%%%%%%%%%%%%%%%%%%%%%%%%%%%%%%%%%%%%%%%%%%%%%%%%%%%%%%
%%%%% References %%%%%

\bibliography{spie}   %>>>> bibliography data in report.bib
\bibliographystyle{spiebib}   %>>>> makes bibtex use spiebib.bst

\end{document}